\documentclass[12pt]{article}
\usepackage{amssymb}
\usepackage{amsmath}
\usepackage{amsthm}
\usepackage{latexsym}
\usepackage{amsfonts}
\usepackage{subeqnarray}
\usepackage{color,graphicx}
\usepackage{graphics}
\usepackage[cp1251]{inputenc}
\inputencoding{cp1251}
\usepackage[T2B]{fontenc}
\usepackage[russian,english]{babel}
\theoremstyle{plain}

\newcommand{\nc}{\newcommand}

\nc{\half}{\frac{1}{2}}
\nc{\reff}[1]{(\ref{#1})}% Put parentheses around equation references
\nc{\ts}{\textstyle}
\nc{\ds}{\displaystyle}
\nc{\nn}{\nonumber}
\def\sop{\left\{p_{k}\right\}_{k=0}^{\infty}}

\def\sAn{\left\{A_{n}^{(j))}\right\}_{n=0}^{\infty}}
\begin{document}

{\bf V.V. Borzov \footnote{The St.-Petersburg State University of Telecommunications, St.-Petersburg,
Russia. E-mail: borzov.vadim@yandex.ru},
E.V. Damaskinsky \footnote{Military institute (technical engineering), St.-Petersburg,
Russia. E-mail: evd@pdmi.ras.ru}}
\bigskip

\centerline{\large\bf Orthogonal polynomials}
\bigskip

\centerline{\large\bf and deformed oscillators}
\bigskip

\begin{quote}
We discuss the construction of oscillator-like systems associated with
orthogonal polynomials on the example of the Fibonacci oscillator.
In addition, we consider the dimension of the corresponding lie algebras.
\end{quote}
\bigskip

\bigskip

{\bf Introduction.}
The basis for this work was the talk given by one of the authors (EVD) at the conference
"In search of Fundamental Symmetries" dedicated to the 90th anniversary Yuri V. Novozhilov.
Therefore, the writing style is not quite academic in nature.

The authors studied in the physics faculty  of Leningrad State University during the period
when the department of "Field Theory and Elementary Particle Physics" was headed by Yu. V. Novozhilov,
who supported the education traditions formed under the influence of academicians
V.I. Smirnov and V.A. Fock.
High level of physical and mathematical education in the physical department was opened
for its graduates wide possibilities of application of the forces
both in physics and in mathematics.
This circumstance explains the fact that the author  who received physical education (EVD) became
interested in the achievements of mathematical physics, and the author  who received the
mathematical education (VVB) became interested in physical problems.
As a result, since 2000, the authors have combined their efforts in the study of
algebraic structures of quantum physics.
In this paper we present some results obtained by the authors in recent years.
These results are associated with the generalization of the notion of a harmonic oscillator, which is one
of the basic concepts of quantum mechanics.

The development of quantum physics in the last century, in particular,
the emergence of quantum groups and quantum algebras
\cite{01}-\cite{04} (L.D. Faddeev school, V. Drinfeld [1983/87]),
naturally led to construction of various generalizations of the notion of
quantum harmonic oscillator\cite{11}-\cite{14}, (L.C. Biedenharn, J.A. Macfarlane,
P.P. Kulish and E.V. Damaskinsky [1989/90])
connected with q-deformations of the canonical
commutation relations of the algebra of harmonic oscillator \footnote{Although several
attempts of generalization of the Heisenberg commutation relations
were made before \cite{05}-\cite{10}.}.
Further researches led to the construction of generalizations of the Heisenberg algebra associated
with orthogonal polynomials from the Askey - Wilson scheme and their q-analogues \cite{16}.

In the work \cite{25} one of the authors (VVB) developed a general scheme for the construction of
such oscillator - like algebras for an arbitrary family of orthogonal polynomials on the real axis
(hereinafter such algebras are called the generalized oscillators).
In recent years we have applied this approach to the construction  of generalized
oscillator algebras associated with
\begin{itemize}
\item some classical orthogonal polynomials of a continuous argument
(Laguerre, Chebyshev (first and second kind), Legendre, Gegenbauer and Jacobi  polynomials);
\item some classical orthogonal polynomials of a discrete argument
(Meixner, Charlier, Kravchuk  polynomials);
\item some q-analogues of the classical orthogonal polynomials (discrete and continuous
q-Hermite polynomials, q-Charlier polynomials);
\item generalized Fibonacci polynomials;
\item the Chebyshev - Koornwinder polynomials of two variables.
\end{itemize}
In addition, in some cases
were  constructed and investigated the corresponding
coherent states of Barut - Girardello and Klauder - Gaseau type.
\bigskip

{\bf The construction of a generalized oscillator associated with the system
of orthogonal polynomials.}
Let we have the system of orthogonal polynomials $\sop$, forming an orthogonal basis
in Hilbert space  $\mathcal{H}_{\mu}=\mathrm{L}^2(\mathbb{R};\mu(dx))$.
For a given measure $\mu$, one can define oscillator-like system for which these
polynomials play the same role as the Hermite polynomials for the standard quantum harmonic oscillator.
Here we briefly describe some details of this construction (for a detailed description see \cite{25}).

Now, let $\mu$ be a (symmetric) probability measure on ${\mathbb R}$ with finite moments
$\mu_n=\int_{-\infty}^{\infty} x^n\,d\mu$ ($\mu_{2k+1}=0$).
These moments uniquely define a positive sequence
$\left\{b_{n}\right\}_{n=0}^{\infty}$
and a system of orthogonal polynomials with recurrent relations
\begin{equation}\label{1}
x\,p_{n}(x)=b_{n}\,p_{n+1}(x)+b_{n-1}p_{n-1}(x).
\end{equation}
These polynomials form an orthogonal basis in the Hilbert space $\mathcal{H}_{\mu}$.
The relations \reff{1} determine the action of the operator coordinates  $X_{\mu}$
$$
X_{\mu}\,p_{n}(x)=b_{n}\,p_{n+1}(x)+b_{n-1}p_{n-1}(x).
$$
Similarly, one can define the momentum operator $P_{\mu}$, which is conjugate
to the operator coordinates with respect to the basis
$$
P_{\mu}\,p_{n}(x)=i(-b_{n}\,p_{n+1}(x)+b_{n-1}p_{n-1}(x)).
$$
Then one can define creation and annihilation  operators
(conjugate to each other in the Hilbert space ${\mathcal H}_{\mu}$)
$a^{\pm}_{\mu}=\frac{1}{\sqrt{2}}\left(X_{\mu}\pm iP_{\mu}\right),$
operator $N_{\mu},$ numbering basis states, and selfadjoint Hamiltonian
$H_{\mu}=X_{\mu}^2+P_{\mu}^2=a^{+}_{\mu}a^{-}_{\mu}+a^{-}_{\mu}a^{+}_{\mu},$ moreover
$$
a^{+}_{\mu}p_{n}(x)=\sqrt{2}b_{n}p_{n+1}(x);\,\,
a^{-}_{\mu}p_{n}(x)=\sqrt{2}b_{n-1}p_{n-1}(x);
$$
$$
N_{\mu}p_{n}(x)=np_{n}(x),\quad
H_{\mu}p_{n}(x)=\lambda_np_{n}(x),
$$
where
$\lambda_0=2b_{0}^2,\quad \lambda_n=2(b_{n-1}^2+b_{n}^2).$
These operators satisfy the commutation relations of
generalized Heisenberg algebra
$$
\left[ a^{-}_{\mu}, a^{+}_{\mu} \right]=
2\left( B(N_{\mu}+I)-B(N_{\mu}) \right);\quad
\left[ N_{\mu}, a^{\pm}_{\mu} \right]=\pm a^{\pm}_{\mu},
$$
where the operator-function $B(N_{\mu})$ is defined by the relation
$$
B(N_{\mu})p_{n}(x)=b_{n-1}^2p_{n}(x),\qquad p_{n}(x)\in{\mathcal H}_{\mu}.
$$
The center of this algebra is generated by the element
${\mathcal C}=2B(N_{\mu})-a^{+}_{\mu}a^{-}_{\mu}.$
With appropriate changes (see \cite{25}) the same reasoning applies to
a more general case (with  asymmetric measure  $\mu$ )
\begin{equation}\label{2}
x\,p_{n}(x)=b_{n}\,p_{n+1}(x)+a_{n}\,p_{n+1}(x)+b_{n-1}p_{n-1}(x).
\end{equation}

Coherent states of Barut - Girardello type for such
generalized oscillator are determined by the relations
$$
a^{-}|z>=z|z>,\quad |z>={\mathcal N}^{-1/2}(|z|^2)\sum_{n=0}^{\infty}
\frac{z^n}{\left( \sqrt{2} b_{n-1}\right)!}p_{n},
$$
where
$$
{\mathcal N}(|z|^2)=<z|z>=\sum_{n=0}^{\infty}
\frac{|z|^{2n}}{\left( 2 b_{n-1}^2\right)!}.
$$
It is possible to prove that the so-defined coherent states form
an overcomplete family of states in the Hilbert space. In addition,
these states  minimize the corresponding uncertainty relation.

The main difficulty in applying this approach is the solution
the  moment  problem that arises when constructing measures of orthogonality
for polynomials (not in the classical case) and in finding of measures, participating in the
(over)completeness relation of the coherent states.
There are also some problems with obtaining the explicit form of coherent states in
terms of special (hypergeometric or basic hypergeometric) functions.
\bigskip
 	
{\bf Fibonacci Numbers and the Fibonacci oscillator.}
In 1202\,, Italian merchant and mathematician Leonardo of Pisa (1180-1240),
known as Fibonacci, published the essay "Liber Abaca".
In this work were collected almost all of the mathematical information  known to that time.
In particular, from this book the European mathematics, using the Latin calculation system,
learned  about Arabic (decimal) one, which is significantly
simplified arithmetic calculations.
Among the many problems given in this book was widely known, "problem of rabbits" \footnote{Some
extensions of this problem, for example, the problem of
mortal rabbits discussed in \cite{26,27}}, the solution of which gives
the sequence of numbers known as the Fibonacci numbers:
$$
F_n: 1,1,2,3,5,8,13,21,34,55,89, \ldots\, n=0,1,2,\ldots .$$
The elements of this sequence (Fibonacci numbers) are determined by the recurrence relation
$$
F_n=F_{n-1}+F_{n-2}\quad (n\geq 2)
$$
with initial conditions $F_0=1,\, F_1=1$.  \footnote{Elementary properties of Fibonacci numbers
are given in a brochure by N.N.Vorobiev \cite{28}.}

It is known \cite{33,34,35} that $F_n$ are associated with the Chebyshev polynomials of the 2nd kind $U_n(x)$ by the relation
$$
F_{n+1}=(-i)^nU_n(i\,{\rm sinh}(\theta_0)), \qquad
U_n(\cos\theta)=
\frac{\sin((n+1)\theta)}{\sin\theta},
$$
where $\theta_0>0$ и ${\rm sinh}(\theta_0)=\half.$

There are many generalizations of the Fibonacci numbers. The most natural of them is defined by the relation
\begin{equation*}
{\mathfrak F}_{n+1}^{a,b}=a\,{\mathfrak F}_{n}^{a,b}+
b\,{\mathfrak F}_{n-1}^{a,b}.
\end{equation*}

Using the relationship of the Fibonacci numbers with Chebyshev polynomials,
Ismail \cite{35} suggest their one-parametric generalization
\begin{equation*}
F_{n+1}(\theta) = (-i)^n\;  U_n(\sinh(i\theta)),\quad
 F_n(\theta) = e^{(n-1)\theta}\; \frac{1-Q^n}{1-Q}
\end{equation*}
where $Q= - e^{-2\theta}.$  These generalizations satisfy the recurrent relations
\begin{equation*}
y_{n+1}(\theta) = 2\sinh \theta y_n(\theta) + y_{n-1}(\theta)
\end{equation*}
with initial conditions
$F_1(\theta)=1,\, F_2(\theta)=2\sinh\theta$.
Ismail showed that $F_n(\theta)$ generate the classical moment problem for measure
\begin{equation*}
\nu(x) =
(1-q^\alpha)\sum_{k=0}^\infty q^{\alpha k}\delta(x- q^ke^{-\theta}),
\end{equation*}
where $\delta(x-c)$ is unique discrete measure concentrated at the point
$x=c$. When $\alpha$ is even this measure is positive with unit total mass,
and in other cases, $\nu$ is the unit "signed" measure.
This measure correspond to orthogonal polynomials  $\{p_n(xe^\theta; q^{\alpha -1}, 1) \}$,
known as the little $q$-Jacobi polynomials, which explicit form looks as
$$
\begin{gathered}
p_n(x; a, b) = {}_2\phi_1(q^{-n}, abq^{n+1}; aq;q,qx) \\
= \sum_{j=0}^n \frac{(q;q)_n(abq^{n+1};q)_j}{(q;q)_j(q;q)_{n-j}}
q^{\binom{j+1}{2}} \frac{(-x)^j}{(aq;q)_j},
\end{gathered}
$$
where $q$-factorials are defined by the formula
$$(\lambda;q)_s = (1-\lambda)(1-\lambda q) \dots (1-\lambda q^{s-1}).$$

Richardson \cite{54} noted that the matrix ${\mathfrak F}_n$, with elements
$\frac{1}{F_{i+j+1}},$ where $F_n$ --- $n$-th Fibonacci number, has as its inverse the matrix with
integer elements.
(Since the same property has the Hilbert matrix, he called this matrix as
the Filbert matrix; this term was established according to the rule Fi(bonacci)+(Hi)lbert).
Using this result, Berg \cite{55} showed that the sequence $\ds\frac{1}{F_{n+2}}$ of numbers inverse to
Fibonacci numbers are a sequence of moments for discrete probability measure.
He also found that this measure is an orthogonality measure for small q-Jacobi polynomials
$$
p_n(x;a,b;q)={}_2\phi_1\left(\genfrac{}{}{}{}{q^{-n},abq^{n+1}}{aq};q,xq\right),
$$
for $a=q,\,b=1$ and $q=\ds\frac{1-\sqrt{5}}{1+\sqrt{5}}$.

Applying the described above method of constructing of the generalized oscillator
to the case of polynomials $p_n(x)\equiv p_n(x;a,b;q)$ we get
appropriate oscillator-like system and a set of coherent
states for her. Further, this system is called the Fibonacci oscillator.
\footnote{Note that in the literature (following \cite{56}) by oscillator Fibonacci sometimes means
two-parameter deformed oscillator \cite{51},
associated with the basic number of $[n;q,p]=\ds\frac{q^n-p^{-n}}{q-p^{-1}},$
which (as well as other basic numbers) satisfies the generalized variant of Fibonacci recurrent relations
$$
[n+1;q,p]=(q+p)^{-1}[n;q,p]-qp^{-1}[n-1;q,p],
\quad [0;p,q]=0,\quad [1;p,q]=1.
$$
}

In our case, recurrence relations have the form
$$
-xp_{n}(x)=A_n p_{n+1}(x)-(A_n+C_n) p_{n}(x) + C_n p_{n-1}(x),
$$
where $p_{0}(x)=1$ and
$$
A_n= q^n \, \frac{(1-aq^{n+1})\,(1-abq^{n+1})}{(1-abq^{2n})\,
(1-abq^{2(n+1)})},
C_n= aq^n \, \frac{(1-q^{n})\,(1-bq^{n})}{(1-abq^{2n})\,(1-abq^{2n+1})}.
$$

Let us denote $p_{n}(x)=\gamma_{n}\Psi_{n}(x),$ where 
$$
\gamma_{n}=\sqrt{\frac{C_{1}C_{2}\cdots C_{n}}{A_{0}A_{1}\cdots A_{n-1}}}=
\left( a^nq^n\frac{1-abq}{1-abq^{2n+1}}
\frac{(q,q)_{n}(bq,q)_{n}}{(aq,q)_{n}(abq,q)_{n}}\right)^{1/2}.
$$
Note that in the case $a=q,\,b=1$ we have
$$
\gamma_{n}=
q^n\frac{(q,q)_{n}}{(q^2,q)_{n}}\sqrt{\frac{1-q^2}{1-q^{2(n+1)}}}.
$$

Then for $\Psi_{n}(x)$ we obtain
$$
x\Psi_{n}(x)=-b_{n} \Psi_{n+1}(x)+a_n \Psi_{n}(x) - b_n \Psi_{n-1}(x),
$$
where $\Psi_{0}(x)=1,$  $a_n=A_{n}+C_{n}$ and
$b_{n-1}=\sqrt{A_{n-1}C_{n}}.$

For $a=q,\,b=1$  we have
$$
a_n=\frac{q^n}{1-q^{2(n+1)}}\left(\frac{(1-q^n)^2}{1-q^{2n+1}}+
\frac{(1-q^{n+2})^2}{1-q^{2n+3}}\right),
$$
$$
b_{n-1}=\frac{q^n}{1-q^{2n+1}}\,
\frac{(1-q^n)(1-q^{n+1})}{\sqrt{(1-q^{2n})(1-q^{2(n+1)})}}.
$$

Now let
$
X_{\mu}=\text{Re}(\widetilde{X}_{\mu}-\widetilde{P}_{\mu}),\,
P_{\mu}=(-i)\text{Im}(\widetilde{X}_{\mu}-\widetilde{P}_{\mu}),
$
where operators $\widetilde{X}_{\mu}$ and $\widetilde{P}_{\mu}$
are defined as
$$
\widetilde{X}_{\mu}\Psi_{n}=b_{n-1}\Psi_{n-1}+a_n\Psi_{n}+b_n\Psi_{n+1};
$$
$$
\widetilde{P}_{\mu}\Psi_{n}=
i\left(b_{n-1}\Psi_{n-1}+a_n\Psi_{n}-b_n\Psi_{n+1}\right).
$$

As a result, using the above relations, we define the
algebra of Fibonacci  oscillator \footnote{Note that in the papers \cite{57}-\cite{58}
also discussed deformed oscillators associated with (generalized)
Fibonacci numbers.}.

Coherent states of Barut - Girardello type  for this oscillator are
defined as above by the relations
$$
a^{-}|z>=z|z>,\quad |z>={\mathcal N}^{-1/2}(|z|^2)\sum_{n=0}^{\infty}
\frac{z^n}{\left( \sqrt{2} b_{n-1}\right)!}\Psi_{n}.
$$
 After some transformations we obtain the following expression
for normalization factor
$$
{\mathcal N}(|z|^2)={}_6\phi_1\left(
\genfrac{}{}{0pt}{}{-q,-q^{3/2},q^{3/2},q^{3/2},-q^{3/2},-q^2}{q^2}
\biggl|\,q;\,\frac{|z|^2}{2}\right).
$$
where by definition
$$
{}_6\phi_1\!\left(\!\genfrac{}{}{0pt}{}{a_1,a_2,a_3,a_4,a_5,a_6}{b_1}
\biggl|\,q;z \!\right)=\!\!\\
\sum_{k=0}^{\infty}
\frac{(a_1,a_2,a_3,a_4,a_5,a_6;q)_k}{(b_1;q)_k}(-1)^{2k}q^{-2k\binom{k}{2}}.
$$
Here we used the standard notation for the Pochhammer $q$-symbol:
$$
(a;q)_0=1,\quad (a;q)_n=(1-a)(1-aq)\cdots(1-aq^{n-1}),
$$
$$
(a_1,a_2,\ldots,a_m;q)_n=(a_1;q)_n(a_2;q)_n\cdots(a_m;q)_n.
$$

In the case $a=q,\,b=1$, the expression for the coherent states is simplified
and has the form
$$
|z>={\mathcal N}^{-1/2}(|z|^2)\sum_{n=0}^{\infty}
q^{-n(n+3)/2}\,p_n(x;q,1|q)\,\frac{(q^3;q)_{2n}}{(q;q)_n^2}\,
\left(\frac{z}{\sqrt{2}}\right)^n,
$$
with the same expression for the normalizing factor. To complete
our construction it would be desirable to find an explicit expression for the measure participating in the
overcompleteness relation for coherent states of the Fibonacci oscillator.
 Currently, this task is still under development.
\bigskip

{\bf The dimension of generalized oscillator algebras}.
In a recently published paper \cite{59} the authors investigated the conditions under which
algebra of the generalized oscillator $\mathfrak{A}$, associated with orthogonal polynomials in the manner described above,
is finite-dimensional. In \cite{59} considered {\bf only} case of orthogonal polynomials
for a symmetric measure on the real axis (when the Jacobi matrix corresponding to the recurrent
relations \reff{1} has zero diagonal). In the work \cite{59a} we have clarified the formulation of sufficient conditions
and extended the results to the case of the Jacobian matrix with nonzero diagonal, corresponding to recurrent
relations \reff{2}. Following property holds \cite{59a}:

{\bf Теорема}. Let us define the sequence
\begin{equation}\label{3}
A_n^{(0)}=b_{n}^{\, 2}-b_{n-1}^{\, 2},\ldots,A_n^{(j)}=A_{n+1}^{(j-1)}-A_n^{(j-1)},
\end{equation}
$j=1,2,\ldots,\quad n=0,1,\ldots$.
Then
\begin{enumerate}
  \item If for any fixed $j>0$, the sequence $\sAn$ is not constant, i.e.
$A_n^{(j)}\neq$const, $n=0,1,\ldots$, then the generalized oscillator algebra $\mathfrak{A}$
is infinite dimensional.
  \item The generalized oscillator algebra $\mathfrak{A}$ is finite dimensional if and only if
\begin{equation}\label{4}
 b_{n}^{\, 2}=(\beta_0+\beta_2n)(1+n),\qquad \beta_0, \beta_2\in\mathbb{R}.
\end{equation}
and in this case dimension of  $\mathfrak{A}$ is equal 4. \hfill $\Box$
\end{enumerate}

Let us consider some examples illustrating this theorem.

As a first example we consider the case of Chebyshev polynomials of the first kind $T_n(x)$
which was not considered in \cite{59}. The polynomials $T_n(x)$ is defined by the relation
\begin{equation*}
T_n(x)=\cos(n\arccos(x)),\qquad x\in[-1,1],
\end{equation*}
and orthogonal in the Hilbert space
$\text{L}^2_{[-1,1]}(\frac{1}{\pi}\sqrt{1-x^2}\text{d}x).$
Normalized polynomials
\begin{equation*}
\Psi^{-\frac{1}{2}}_n(x)\!=\!\sqrt{2}T_n(x),\, n\geq 1,\quad
\Psi^{-\frac{1}{2}}_0(x)\!=\!T_0(x)\!=\!1,
\end{equation*}
fulfill the symmetric recurrence relations \reff{1}  with
\begin{equation*}
b_n=\frac{1}{2},\quad n\geq 1,\qquad b_0=\frac{1}{\sqrt{2}}.
\end{equation*}
In this case $A_n^{(j)}\neq$constant, $\quad n=0,1,\ldots$,
for all fixed $j>0$ and, consequently, the corresponding algebra $\mathfrak{A}$ is infinite dimensional.

As a second example, we consider the Laguerre polynomials satisfying nonsymmetric recurrent relations.
These polynomials
\begin{equation*}
L_n^{\alpha}(x)=\frac{\alpha+1}{n!}{}_1F_1(-n,\alpha+1;x).
\end{equation*}
are orthogonal in the Hilbert space
$\mathcal{H}=\text{L}^2(\mathbb{R}_+^1;x^{\alpha}\exp(-x)\text{d}x)$.
Normalized polynomials
\begin{equation*}
\Psi_n(x)\!=\!d_n^{-1}L_n^{\alpha}(x),\quad d_n\!=
\!\sqrt{\frac{\Gamma(n\!+\!\alpha\!+\!1)}{n!}},\,\, n\geq0
\end{equation*}
fulfill the nonsymmetric recurrence relations  \reff{2} with
\begin{equation*}
b_n\!=\!-\sqrt{(n\!+\!1)(n\!+\!\alpha\!+\!1)},\quad a_n\!=\!2n\!+\!\alpha\!+\!1.
\end{equation*}
In this case, $b_n^{\,\, 2}$ have the form \reff{4} and, hence,the corresponding algebra
$\mathfrak{A}_L$ 4-dimensional but not isomorphic to the algebra of the harmonic oscillator.

Finally, the last example is the oscillator Fibonacci discussed above.
Indeed, in this case, the coefficients $b_n$ are determined by the formulas \reff{3},
as in the first example, $A_n^{(j)}\neq$ constant, $\quad n=0,1,\ldots$, for all fixed $j>0$.
Consequently, the corresponding algebra $\mathfrak{A}$ is infinite dimensional.
\bigskip

{\bf Acknowledgements.}\, The authors are grateful to P.P. Kulish and V.D. Lyakhovsky for discussion
of a number of issues related to the subject of this article.
Research of EVD performed with the financial support of RFBR, grant 15-01-03148-а.


\begin{thebibliography}{99}
\bibitem{01} P.P. Kulish, N.Yu. Reshetikhin, {\it Quantum linear problem for the  sine - Gordon equation
and higher representations},  Journal of Soviet Mathematics,  {\bf 23}(4), 2435-2441 (1983).
\bibitem{02} L.D. Faddeev, L.A. Takhtajan, {\it Liouville model on the lattice},
 pp.166-179 In: \textit{Field theory, quantum gravity and strings}, Lecture notes in physics {\bf 246}
 Springer 1986.
\bibitem{03} L.D. Faddeev, N.Yu. Reshetikhin, L.A. Takhtajan, {\it Quantization of
Lie groups and Lie algebras}, Leningrad Mathematical Journal,  {\bf 1}(1), 193–225	(1990).
\bibitem{04} V.G. Drinfeld, {\it Quantum groups}, Proc. ICM-86 (Berkeley) {\bf 1}, 798-820 (1987).
\bibitem{11} L.C. Biedenharn, {\it The quantum group $SU_q(2)$ and a q-analogue of the
 boson operators}, J. Phys. A: Math. Gen. {\bf 22}(18), L873-L878 (1989).
\bibitem{12} A.J. Macfarlane, {\it On q-analogues of the quantum harmonic oscillator and
 the quantum group $SU_q(2)$}, J. Phys. A: Math. Gen. {\bf 22}(21), 4581-4588 (1989).
\bibitem{13} P.P. Kulish, E.V. Damaskinsky, {\it On the q-oscillator and the quantum
   algebra $su_q(1,1)$}, J. Phys. A: Math. Gen. {\bf 23}(9), L415-L419 (1990).
\bibitem{14} E.V. Damaskinsky, P.P. Kulish, {\it Deformed oscillators and their applications}.
Journal of Soviet Mathematics,  {\bf 62}:5, 2963–2986 (1992).
\bibitem{05} G. Iwata, {\it Transformation Eunctions in the Complex Domain},
Prog. Theor. Physics, {\bf 6}, No.4, 524 July-August, 1951
\bibitem{06} V.V. Kuryshkin, {\it On some generalisation of creation and annihilation operators in quantum theory
(pquantisation)}, Dep. VINITI Dep. 3936-76 {1976} (in Russian);  Ann. Found. L de Broglie {\bf 5}, 111 (1980).
\bibitem{07} Cigler, J.: {\it Operatormethoden fur q-ldentitaten}, Mh. Math. {\bf 88}, 87-105 (1979).
\bibitem{09} A. Jannussis, G. Bbodimas, D. Sourlas, L. Papaloucas, P. Siafaricas and K. Vlachos,
{\it Some properties of q-analysis and applications to non-canonical mechanics}, Hadronic J. {\bf 6}, 1653-1686 (1983).
\bibitem{10} M. Arik, D.D. Coon, {\it Hilbert spaces of analytic functions and generalized coherent states},
 J. Math. Phys. {\bf 17}(4), 524-527 (1976).
\bibitem{16} Roelof Koekoek, Peter A. Lesky, Rene F. Swarttouw, {\it Hypergeometric Orthogonal Polynomials and Their q-Analogues}, Springer-Verlag Berlin Heidelberg, 2010.
\bibitem{25} V.V. Borzov, {\it Orthogonal polynomials and generalized oscillator algebras},
    Integral Transf. and Special Functions, {\bf 12}(2), 115-138 (2001).
\bibitem{26} V.E. Hoggat, D.A. Lindt, {\it The daying rabbit problem},
 Fibonacci Quart.  {\bf 7:4}, 482-487 (1969).
\bibitem{27} A.M. Oller, {\it The dying rabbit problem revisited}. arXiv\,:\,math/07102216.
\bibitem{28} N.N. Vorobiev, {\it Fibonacci numbers} 1964 (in Russian).
\bibitem{33} R. Askey, {\it Fibonacci and Related Sequences}. Math. Teacher, {\bf 97}:2, 116-119 (2004).
\bibitem{34} R. Askey, {\it Fibonacci and Lucas Numbers}. {\it Math. Teacher} {\bf 98}:9, 610-614 (2005).
\bibitem{35} M.E.H. Ismail, {\it One parameter Generalizations of the Fibonacci
 and Lucas polynomials},  arXiv\,:\,math/0606743 (2006).
\bibitem{50} E.V. Damaskinsky, V.V. Borzov, {\it Fibonacci oscillator}. pp.54-65 In {\it Quantum theory and cosmology}.
 2009 (in Russian).
\bibitem{51} R. Chakrabarti, R. Jagannathan, {\it A $(p,q)$-oscillator realization
 of two-parameter quantum algebras}, J. Phys. A. {\bf 24}:13,  L711-718 (1991).
\bibitem{54} T.M. Richardson, {\it The Filbert matrix}. Fibonacci Quart. {\bf 39}:3, 268-275 (2001).
\bibitem{55} C. Berg, {\it Fibonacci numbers and orthogonal polynomials},\hfill\break arXiv\,: math/0602283.
\bibitem{56} M. Arik, E. Demircan, T. Turgut, L. Ekinci and M. Mungan,
 {\it Fibonacci Oscillators}, Z. Phys. C. {\bf 55}:1, 89-96 (1992).
\bibitem{57} J. de Souza, E.M.F. Curado and M.A. Rego-Monteiro, {\it Generalized Heisenberg Algebras and Fibonacci Series},
 J. Phys. A.  {\bf 39}:33, 10415-10425 (2006).
\bibitem{58} M. Schork, {\it Generalized Heisenberg algebras and $k$-generalized Fibonacci numbers},
J. Phys. A. {\bf 40}:15, 4207-4214 (2007).
\bibitem{59} G. Honnouvo, K. Thirulogasanthar, {\it On the dimensions of the oscillator algebras
induced by orthogonal polynomials}, J. Math. Phys. {\bf 55} , 093511 (2014); arXiv:1305.2509.
\bibitem{59a} V.V. Borzov, E.V. Damaskinsky, {\it On  dimensions of oscillator algebras},
Proc. of {\it Days on Difraction 2014} (edited by IEEE) pp.48-52 (2014).
\end{thebibliography}
\end{document}